**This is the accepted manuscript (postprint) of the following article:**
N. Namvar, E. Salahinejad, A. Saberi, M.J. Baghjeghaz, L. Tayebi, D. Vashaee, *Toward reducing the formation temperature of diopside via wet-chemical synthesis routes using chloride precursors*, Ceramics International, 43 (2017) 13781-13785.
https://doi.org/10.1016/j.ceramint.2017.07.094

# Toward reducing the formation temperature of diopside via wet-chemical synthesis routes using chloride precursors

N. Namvar [a], E. Salahinejad [a,*], A.H. Saberi [a], M. Jafari Baghjeghaz [a], L. Tayebi [b], D. Vashaee [c]

[a] Faculty of Materials Science and Engineering, K.N. Toosi University of Technology, Tehran, Iran

[b] Department of Developmental Sciences, Marquette University School of Dentistry, Milwaukee, WI 53233, USA

[c] Electrical and Computer Engineering Department, North Carolina State University, Raleigh, NC, 27606, USA

## Abstract

Reducing the formation temperature of single-phase multioxides is one of the classic challenges in ceramic processing, including wet-chemical synthesis routes. Toward pursuing this aim for diopside ($MgCaSi_2O_6$), the merit of different sol-gel and coprecipitation processes using the related chloride precursors followed by calcination was compared from the viewpoints of crystallinity and homogeneity. In accordance to the results, the use of the sol-gel techniques, directed with/without an alkaline catalyst, gave rise to the unfavorable creation of multiphase and low-crystallinity structures. Regarding the coprecipitation methods, the one-step addition of a precipitant agent is accompanied by an indirect low-temperature formation of nano-diopside, while a direct crystallization into this phase was explored in the dropwise condition, albeit with a lower crystallinity. Thus, by employing a suitable synthesis processing, it is feasible to take control of a wide range of nanoparticulate diopside-based structures achieved after a low-temperature calcination.

* Corresponding Author: Email Address: <salahinejad@kntu.ac.ir>

## 1. Introduction

Magnesium-containing silicate materials are a relatively new group of multioxide ceramics developed at different crystallinity levels, including glasses, glass-ceramics and crystals. Diopside (MgO-CaO-2$SiO_2$ or $MgCaSi_2O_6$) is a typical member of this family with a monoclinic crystal structure, density of approximately 3.2 gr/cm$^3$ and melting point of 1391 °C. Due to its suitable mechanical properties and biocompatibility, diopside has been used in various areas like bone tissue engineering scaffolds [1], drug delivery [2], surgery hemostasis [3], sealants for fuel cells [4] and immobilization of radioactive wastes [5].

There are different approaches utilized to produce oxide ceramics, from solid-state to wet methods. Typically, wet-chemical synthesis methods like sol-gel and coprecipitation are promising due to the feasibility of nanostructuration and purity, which is advantageous in most applications. The sol–gel process is known as a versatile method to prepare fine particles with a high sinterability at various forms like dense, porous, fiber and film [6]. Also, coprecipitation generally leads to the creation of mono-dispersed nanoparticles [7]. In most cases, the as-synthesized products of both methods need to be followed with a calcination process to form desirable phases, especially for multioxides. However, their products benefit from a uniform molecular-level mixing, thereby giving occasionally desired structures at comparatively low calcination temperatures and appropriate properties [8-22].

The wet-chemical synthesis of diopside has been reported through sol-gel and coprecipitation methods [23, 24]. In these researches, precursors comprised organic tetraethyl orthosilicate (TEOS), inorganic hydrate calcium and magnesium nitrates were used, which resulted in a well-crystallized single-phase diopside structure at the calcination temperature of 1100 °C. The current work focuses on the sol-gel and coprecipitation synthesis of this multioxide using other precursors (less expensive inorganic chloride ones), toward reducing the formation temperature which is more cost-effective and avoids considerable particle coarsening.

## 2. Materials and methods

With the aim of the diopside synthesis by sol-gel and coprecipitation techniques, magnesium chloride ($MgCl_2$, Merck, >98%), calcium chloride ($CaCl_2$, Merck, >98%), silicon tetrachloride ($SiCl_4$, Merck, >99%), ethanol ($C_2H_5OH$, Merck, >99%) and ammonia solution ($NH_4OH$, Merck, 25%) were used as the starting materials. In accordance to the diopside stoichiometry, an equimolar quantity of $CaCl_2$ and $MgCl_2$ was dissolved in $C_2H_5OH$. Afterwards, the suitable content of $SiCl_4$ was added dropwise to the above solution connected to an ice-water bath, reaching a pH of 0.5. For comparative purposes, the below five procedures were separately conducted on the solution:

I) Stirring was continued for 24 h at the same temperature as the non-catalyzed sol-gel method, which leads to a viscous creamy sol.

II) After the addition of the $NH_4OH$ solution at a molar value equal to that of $SiCl_4$, the solution was stirred for 24 h, where pH did not show any considerable change. This catalyzed sol-gel process is terminated to a white gel.

III) The $NH_4OH$ solution was added dropwise to the solution until pH of 10 was reached, leading to the formation of white precipitates deposited under a viscous creamy liquid. The total level of the precipitant added (N) was recorded and used in the below approaches.

IV) N moles of the $NH_4OH$ solution was immediately added to the solution in a one-step coprecipitation process.

V) After vigorous string of the above solution for 2 h, N moles of the $NH_4OH$ solution was immediately poured in the solution-containing beaker (delayed one-step coprecipitation), giving an apparent feature similar to the procedures III and IV.

The precipitates obtained from the five above-described methods were dried at 120 °C for 3 h, calcined at 700 °C for 2 h and analyzed by X-ray diffraction (XRD) for comparison. Furthermore, the precipitates synthesized by the qualified methods, with the criteria of the most homogeneity and crystallinity, were assessed by simultaneous thermogravimetry (TGA). Also, they were calcined at different temperatures and further studied by XRD, transmission electron microscopy (TEM), field-emission scanning electron microscopy (FESEM) and Brunauer–Emmett–Teller (BET) nitrogen gas adsorption analyses.

## 3. Results and discussion

Fig. 1 shows the XRD patterns of the samples synthesized by the different sol-gel and coprecipitation techniques after calcination at 700 °C. According to the peak analysis by the PANalytical X'Pert HighScore software, the sample synthesized by the non-catalyzed sol-gel (I) method presents a multiphase and low-crystallinity structure (Fig. 1a). As well as the predominant phase of diopside ($MgCaSi_2O_6$, Ref. code: 00-017-0318), a low level of undesirable impurity phases including merwinite ($Ca_3MgSi_2O_8$, Ref. code: 00-025-0161), akermanite ($Ca_2MgSi_2O_7$, Ref. code: 00-035-0592) and monticellite ($CaMgSiO_4$, Ref. code:

00-011-0353) was detected in this sample. However, based on Fig. 1b, as a result of using the alkaline catalysis, the crystalline impurities successfully disappear due to the modification of hydrolysis and condensation sol-gel reactions, whereas crystallinity is still low similar to the non-catalyzed process. The selection of this alkaline catalyst, instead of acids which are more common in sol-gel processing, was based on the significant acidity of the solution. The deteriorous effect of the acid addition to super-acidic solutions has been also reported for sol-gel processing of single-phase forsterite ($Mg_2SiO_4$) through retarding the sol-gel condensation reaction [25]. Anyways, the sample synthesized by the method II possesses a heterogeneous structure of diopside and an amorphous phase. The presence of the amorphous phase, meaning the low crystallinity of the sample, were realized from the low intensity of the crystalline peaks. Considering the stoichiometry of both the loaded precursors and the existing crystalline phase, where both match with diopside, the overall composition of the amorphous phase is speculated to be at the diopside (Ca:Mg:Si = 1:1:2) stoichiometry.

Concerning the coprecipitation methods (Figs. 1 c, 1d and 1e), the following points are notable:

1) Considering the intensity of the crystalline peaks, it is found that the coprecipitation methods develop more crystallinity than the sol-gel processes. That is, on this comparison, the increase of the $NH_4OH$ solution usage improves the homogeneity of the as-synthesized product and thereby the calcined crystallinity. Because by this addition, the acidic solution of the precursors is buffered from pH of 0.5 to 10 during the wet-chemical synthesis routes. This is in good agreement with the fact that the excessive acidity of the solution challenges the synchronous hydrolysis and condensation reactions of precursors used in sol-gel processing of multioxides [25, 26].

2) Apart from diopside, there is no crystalline impurity in the calcined products of the three employed coprecipitation processes. It is another evidence for the improved homogeneity of the as-synthesized products obtained from the coprecipitation processes, in comparison to the sol-gel process (Fig. 1a).

3) The crystallinity of diopside is enhanced from the methods III to V. That is, the sudden addition of the precipitant is preferred to the dropwise addition from the viewpoint of crystallinity. In this regard, the highest crystallinity belongs to the modified coprecipitation (V) approach in which the alkaline precipitant was suddenly added to the acidic precursor solution after the enough homogenization of the solution. Indeed, during the dropwise addition of the precipitant in the approach III, pH is gradually enhanced from 7 to 10 and precipitation occurs in a range of pH. This imposes an inhomogeneity in the as-synthesized product, enhancing the calcination temperature for a homogeneous diopside structure. Whereas the sudden addition of the precipitant to the homogeneous solution (the method V) guarantees the formation of the product at a relatively constant pH with a high uniformity.

The XRD pattern of the samples synthesized by the drop coprecipitation method III after calcination at different temperatures is indicated in Fig. 2. As can been seen, the samples calcined at 500 and 600 °C exhibit a fully amorphous structure. Crystallization starts at 700 °C and is developed at 1000 °C at the phase of diopside. On the other hand, in accordance to the TGA profile of the as-synthesized III product (Fig. 3a), the evaporation and sublimation of the process residuals have been finished before 500 °C. It means that amorphous residual-free diopside can be prepared by the drop coprecipitation method followed by a calcination temperature as small as 500 °C . In comparison, the lowest calcination temperature to develop amorphous residual-free diopside has been reported to be 750 °C in the literature, due to the use of nitrate precursors [23, 24]. In contrast, the chloride residuals have already left the material at about 340 °C which is related to the sublimation point of ammonium chloride formed due to the reaction of chloride (originating from the precursors) with ammonia (coming from the precipitant). This reduced temperature is remarkable because it limits the required energy consumption and particle coarsening.

Fig. 4 presents the XRD patterns of the samples synthesized by the delayed one-step coprecipitation technique V, which presented the most crystallinity among the different methods (Fig. 1), after calcination at different temperatures. The sample dried at 120 °C exhibits a dominant phase of ammonium chloride ($NH_4Cl$, Ref. code: 01-073-0365) and a minor level of ammonium aqua magnesium chloride ($NH_4(Mg(H_2O)_6)Cl_3$, Ref. code: 01-076-1454). Considering the loaded precursors, the dried powder should contain an amorphous phase with the Ca and Si species. After calcination at 350 °C, the amount of $NH_4Cl$ is reduced due to the progress of its sublimation (see the related TGA curve signified in Fig. 3b). In addition, a part of the $NH_4(Mg(H_2O)_6)Cl_3$ and amorphous phases react to create the crystalline phases of akermanite, wollastonite ($CaSiO_3$, Ref. code: 00-027-0088) and magnesia (MgO, Ref. code: 00-004-0829). The phase reactions are continued to 500 °C to dominate magnesia and calcium chloride tetrahydrate ($CaCl_2H_8O_4$, Ref. code: 01-070-0703). The desirable phase of diopside appears at 600 °C, albeit accompanied by silicon dioxide ($SiO_2$, Ref. code: 00-050-1431), enstatite ($MgSiO_3$, Ref. code: 00-019-0768), wollastonite and akermanite, with a low crystallinity. The increase in crystallinity and the diopside amount at 650 °C is noticeable. Eventually, the crystalline phase of diopside dominates at the calcination temperature of 700 °C. In comparison to the drop coprecipitation method demonstrating a direct crystallization into diopside (Fig. 2), an indirect formation of diopside was found for the delayed one-step (V) condition. The increase in the peak intensity of diopside, due to the development of crystallinity and crystallite growth, is also evident by increasing the calcination temperature to 1100 °C.

The need for a calcination temperature of 1100 °C to have diopside has been reported in the literature [23, 24]. It again reflects the merit of the procedures described in the research from the viewpoint of reducing the required temperature from 1100 to 700 °C. This improvement can be attributed to the precursors used in this work, especially $SiCl_4$ which has been rarely used to synthesize multioxides. As a result of adding this precursor into the synthesis solution, pH is reduced to almost 0.5. No precipitation can occur in this significantly high acidic condition; thus, the solution is allowed to become completely homogeneous from the ionic viewpoint before adding the precipitant. In contrast, the Si precursor of TEOS has been normally used in the previous works imposing no considerable change in pH of the solution, which allows minor precipitations even before adding the precipitant. Additionally, by starting the dropwise precipitant addition accompanied by an increase in pH from 7 to 10, precipitation reactions happen in a range of pH. This leads to a compositional inhomogeneity in the as-synthesized product, enhancing the calcination temperature for a homogeneous diopside structure. Whereas the one-step addition of the precipitant to the homogeneous solution developed in this work assures the formation of the product at a relatively constant pH, reducing the required calcination temperature.

The TEM and SEM micrographs of the delayed one-step coprecipitation V derived powder samples after calcination at 120, 500, 700 and 1100 °C are shown in Fig. 5. A statistical analysis on the microscopic images (Fig. 6) reveals that the particle size is enhanced with the calcination temperature, due to particle coarsening via the Ostwald ripening and surface diffusion mechanisms to decrease the total surface energy. However, the rate of this increase is reduced with temperature, which is attributed to a decrease in the driving force of coarsening (i.e. reverse size) with the increase of the size [27]. Also, the BET analysis on the selected powder samples fairly verifies their mean particle sizes, as correlated via the below equation:

$$S^* = \frac{S}{m} = \frac{S}{\rho V} = \frac{\pi r^2}{\frac{4}{3}\pi r^3 \rho} = \frac{6}{\rho D} \quad (1)$$

where $S^*$ is specific surface area (equal to about 12 and 4 $m^2/gr$ for the calcination temperatures of 700 and 1100 °C, respectively), $S$ is surface area, $m$ is mass, $\rho$ is density (equals to 3.27 $gr/cm^3$ for diopside), $r$ is particle radius and $D$ is particle diameter, with the assumption of spherical morphology.

## 4. Summary

In this paper, the wet-chemical synthesis of diopside using the chloride precursors, followed by calcination, was studied. According to the results, the application of the ammonia catalysis improved the homogeneity and crystallinity of the sol-gel derived diopside-based products; while the coprecipitation processes overall developed more desirable features from these viewpoints. The achievement of amorphous residual-free diopside was feasible with the dropwise addition of the ammonia precipitant after calcination at 500 °C. On the contrary, the one-step addition of the precipitant allowed the preparation of crystalline diopside nanoparticles at a temperature as small as 700 °C.

## Figures

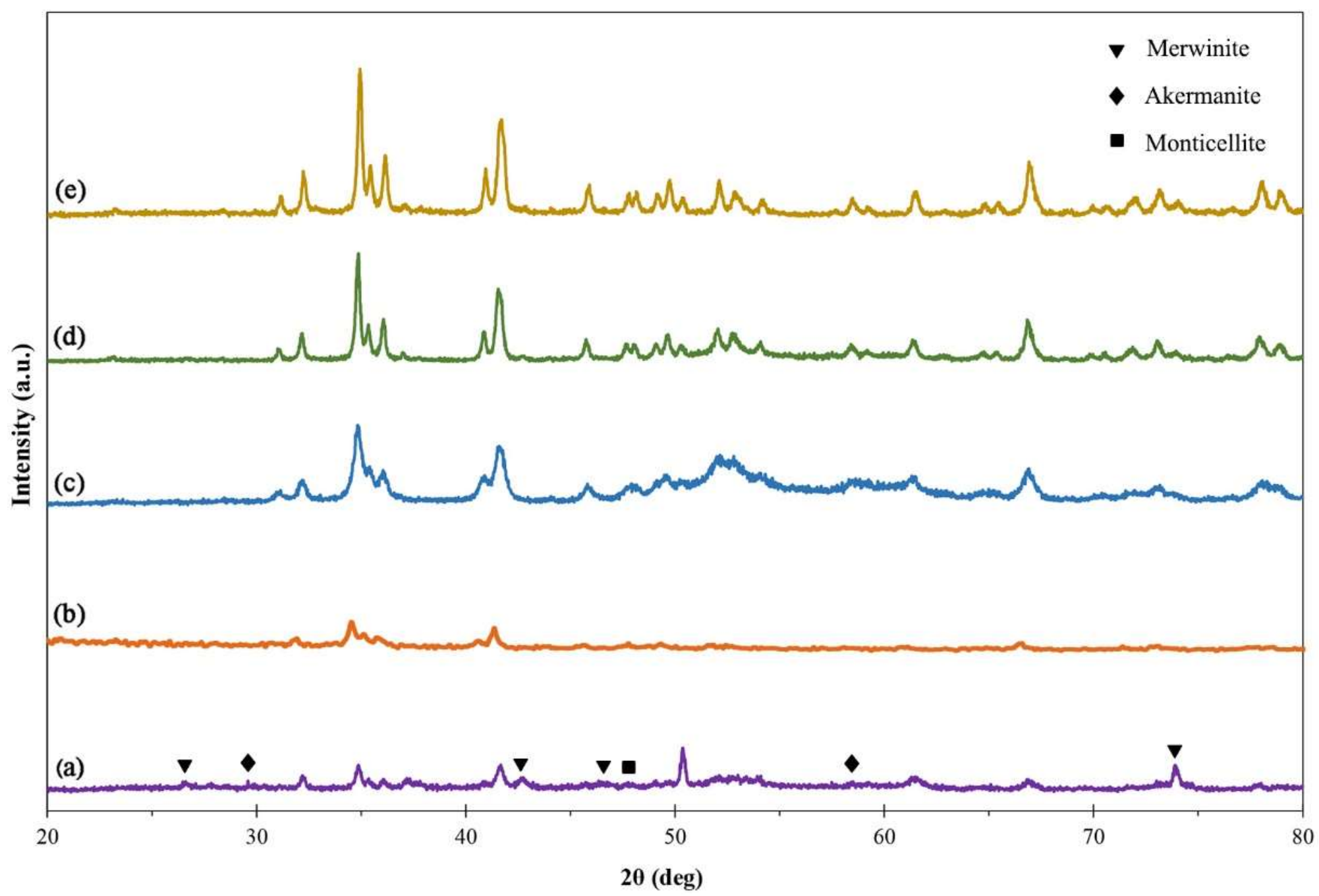

Fig. 1. XRD patterns of the powder samples synthesized by the approaches I (a), II (b), III (c), IV (d) and V (e) after calcination at 700 °C (non-labeled peaks belong to diopside).

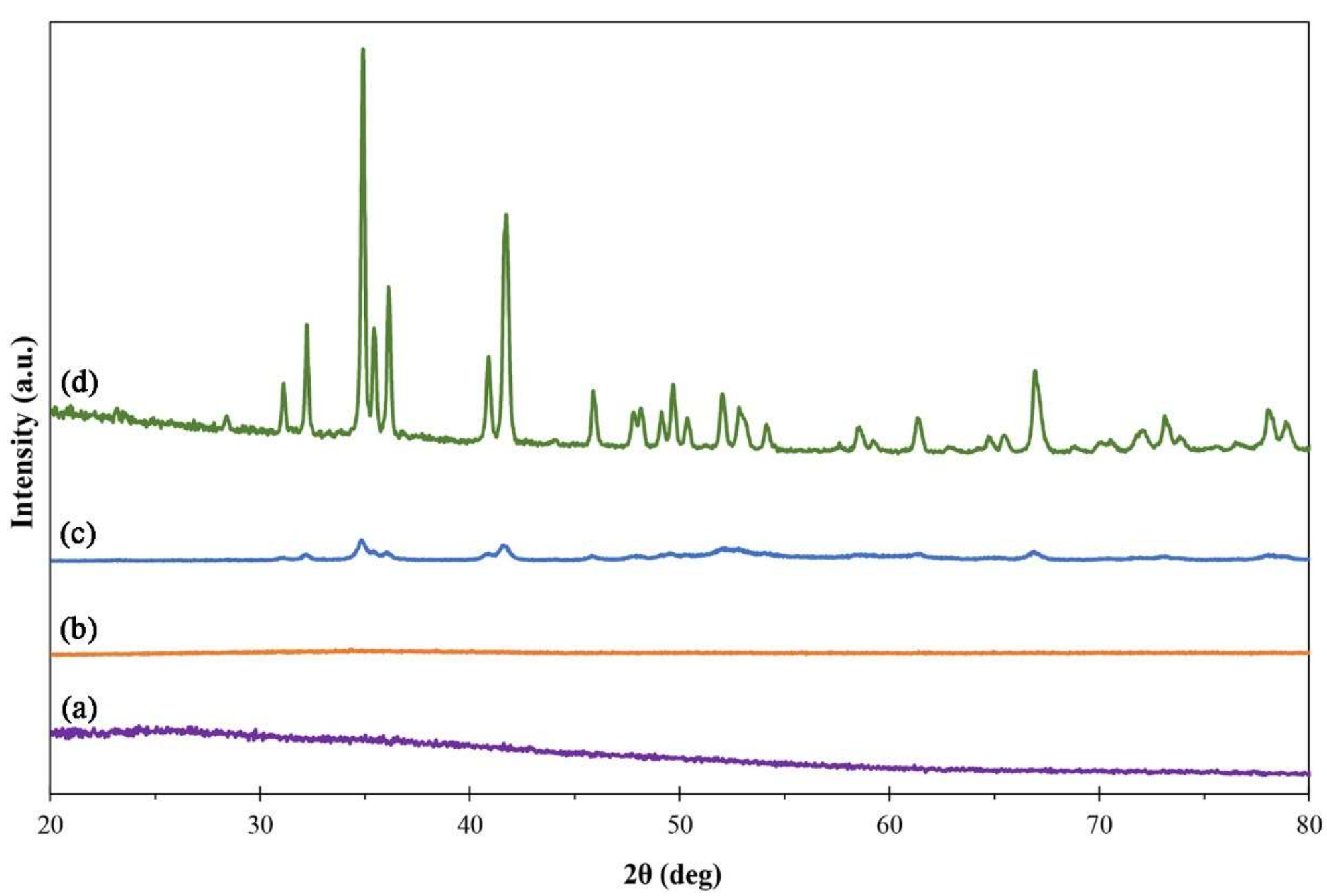

Fig. 2. XRD patterns of the powder samples synthesized by the approach III after calcination at 500 (a), 600 (b), 700 (c) and 1100 (d) °C (non-labeled peaks belong to diopside).

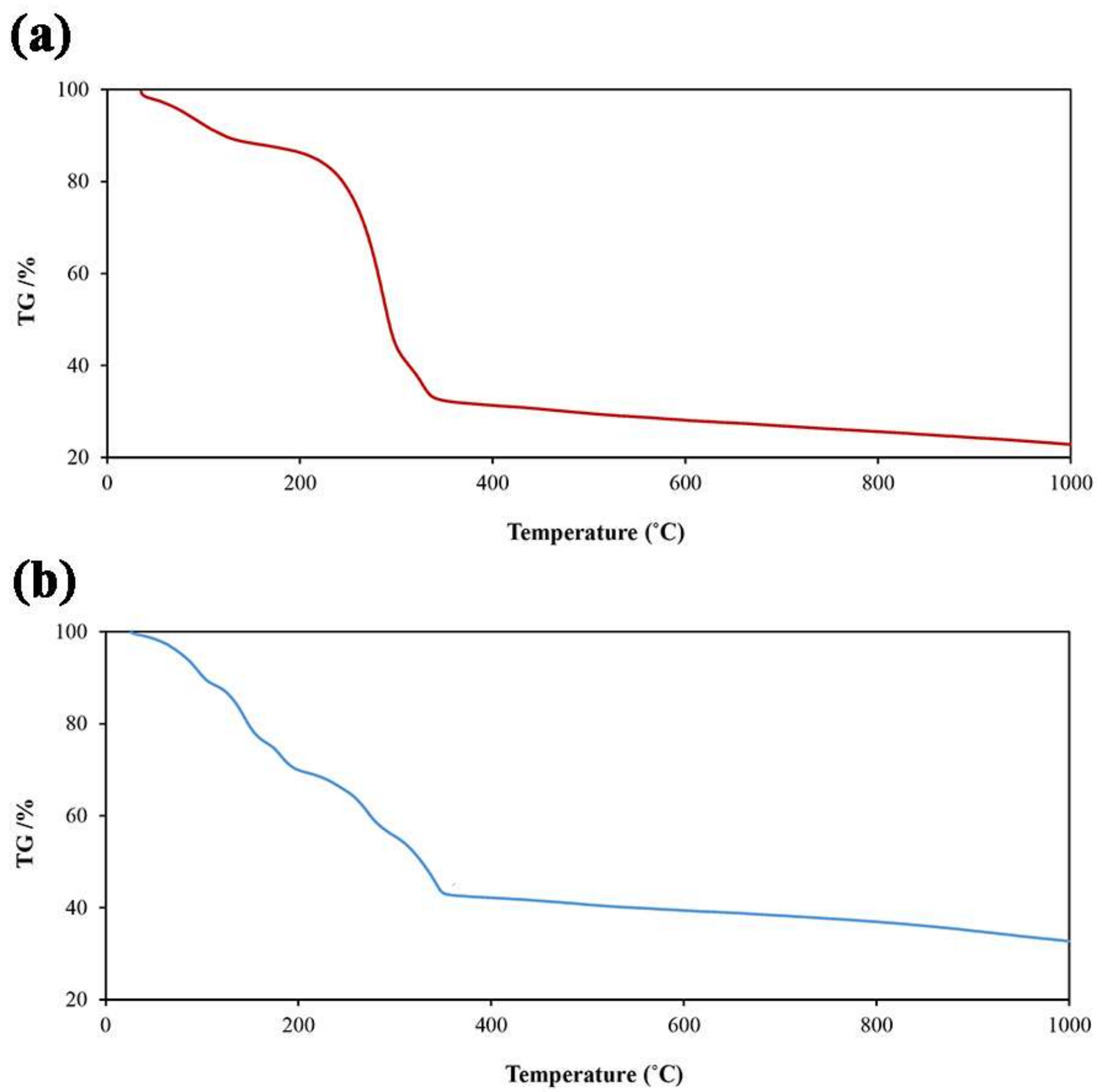

Fig. 3. TGA profiles of the samples synthesized by the approaches III (a) and V (b).

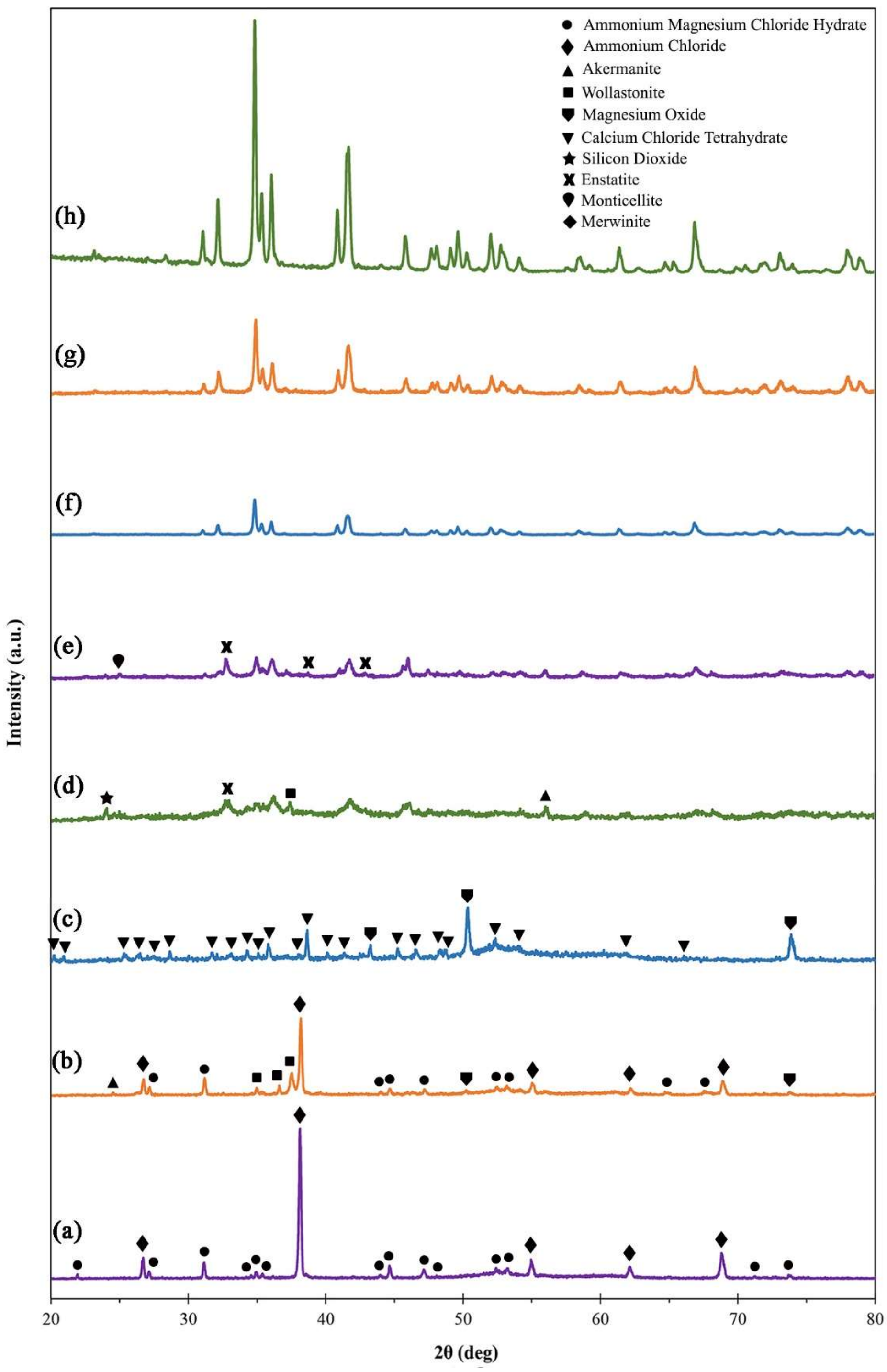
Ammonium Magnesium Chloride Hydrate
Ammonium Chloride
Akermanite
Wollastonite
Magnesium Oxide
Calcium Chloride Tetrahydrate
Silicon Dioxide
Enstatite
Monticellite
Merwinite
(h)
(g)
(f)
(e)
(d)
(c)
(b)
(a)
Intensity (a.u.)
2θ (deg)
20
30
40
50
60
70
80

Fig. 4. XRD pattern of the powder samples synthesized by the approach V after calcination at 120 (a), 350 (b), 500 (c), 600 (d), 650 (e), 700 (f), 800 (g) and 1100 (h) °C (non-labeled peaks belong to diopside).

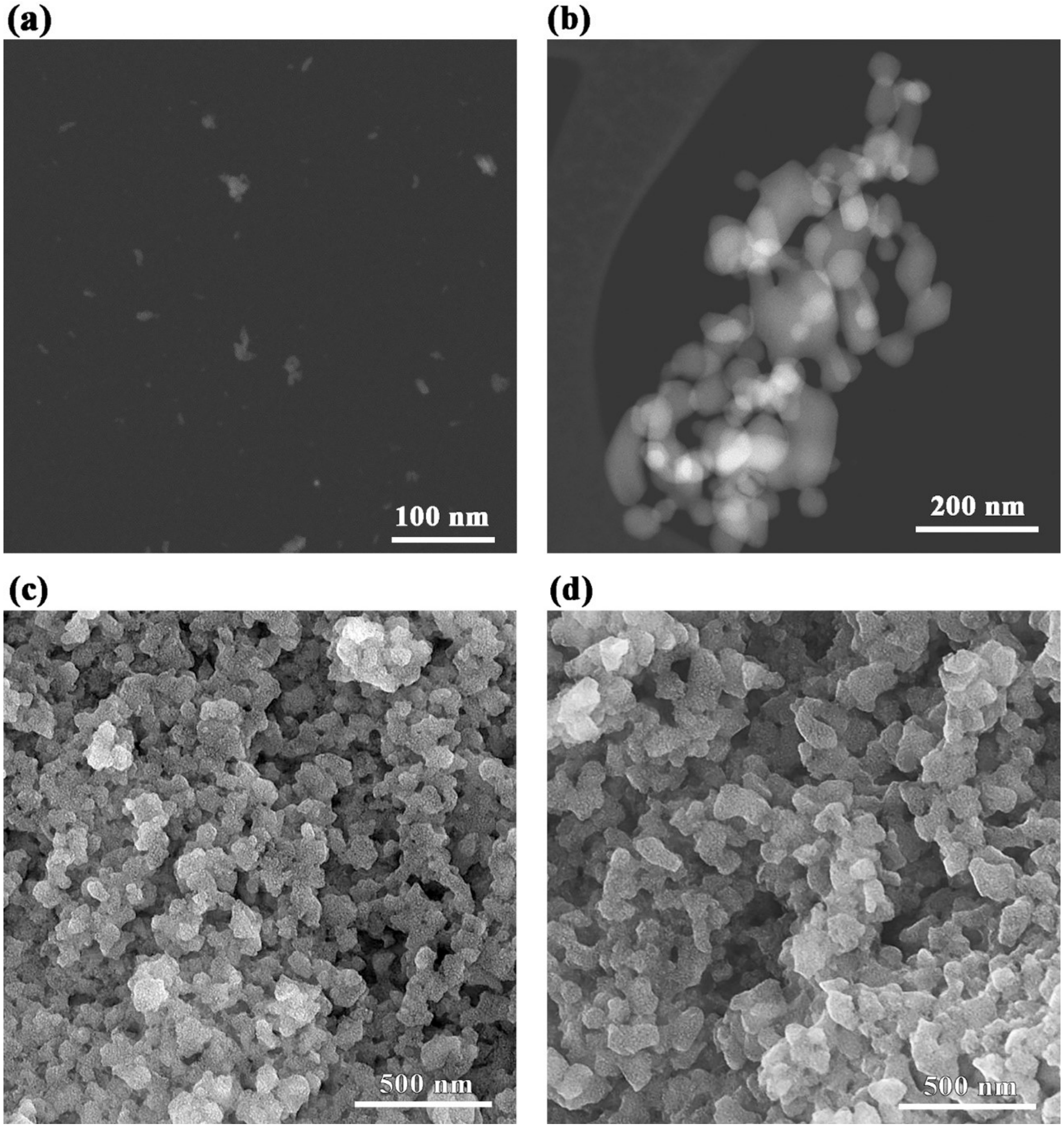

Fig. 5. TEM micrographs of the samples synthesized by the approach V after calcination at 120 (a) and 500 (b) °C, and SEM micrographs of the samples synthesized by the same approach after calcination at 700 (c) and 1100 (d) °C.

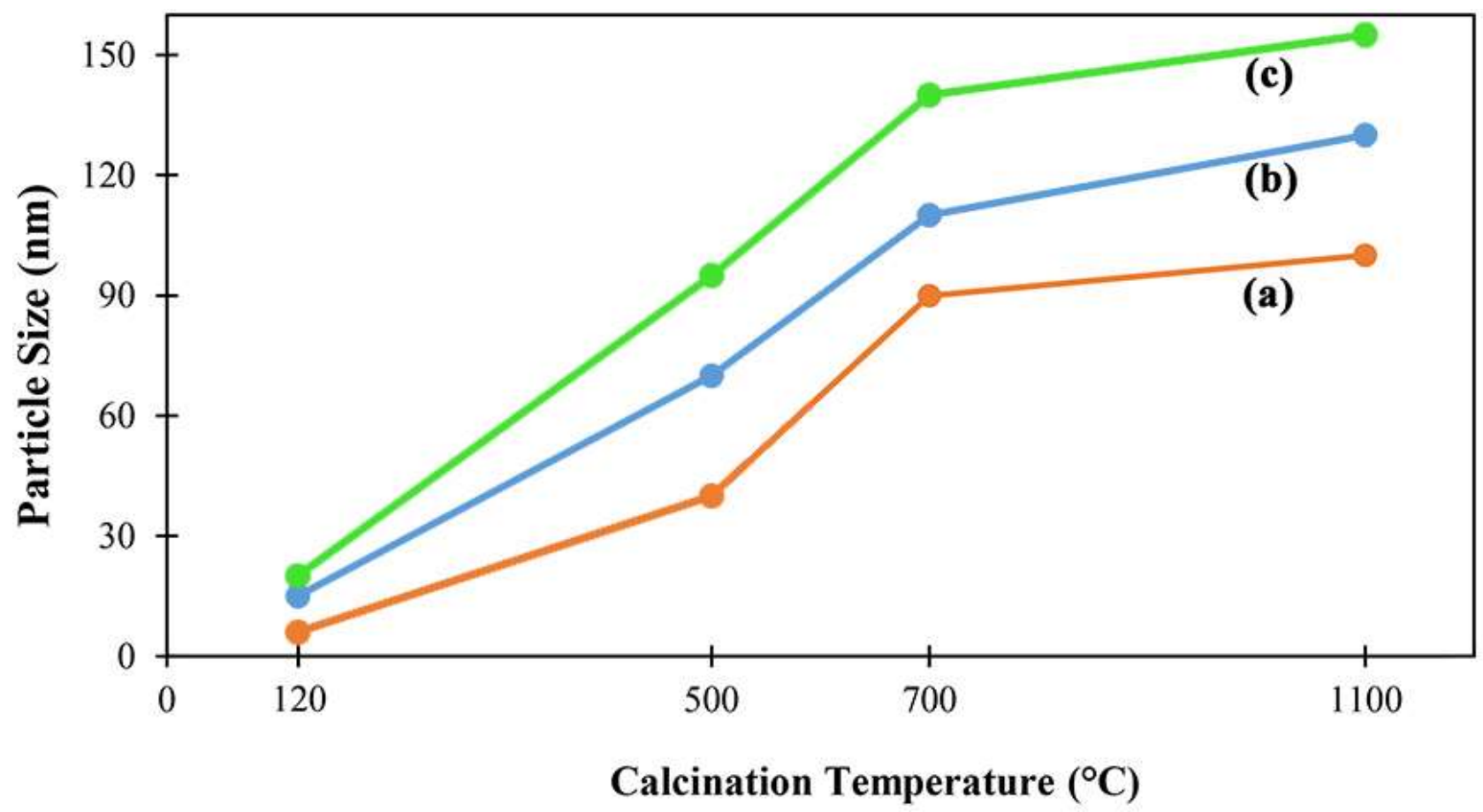

Fig. 6. Smallest (a), arithmetic mean (b) and largest (c) particle sizes of the approach V after calcination at the different temperatures, derived from the micrographs of Fig. 5.